\documentstyle[aps,epsf]{revtex}

\begin{document}
\twocolumn[\hsize\textwidth\columnwidth\hsize\csname@twocolumnfalse\endcsname

\title{$c$-axis  Josephson Tunnelling in Twinned and Untwinned YBCO-Pb Junctions}
\author{C. O'Donovan,\cite{me} M.D. Lumsden, B.D. Gaulin and J.P. Carbotte}
\address{Department of Physics \& Astronomy, McMaster University \\Hamilton, Ontario, Canada L8S 4M1}
\date{July 17, 1996}
\maketitle

\begin{abstract}
Within a microscopic two band model of planes and chains with a
pairing potential in the planes and off diagonal pairing between
planes and chains we find that the chains make the largest
contribution to the Josephson tunnelling current and that through them
the $d$-wave part of the gap contributes to the current.  This is
contrary to the usual assumption that for a $d$-wave tetragonal
superconductor the $c$-axis Josephson current for incoherent
tunnelling into an $s$-wave superconductor is zero while that of a
$d$-wave orthorhombic superconductor with a small $s$-wave component
to its gap it is small but non-zero.  Nevertheless it has been argued
that the effect of twins in YBCO would lead to cancellation between
pairs of twins and so the observation of a current in $c$-axis YBCO-Pb
experiments is evidence against a $d$-wave type order parameter. We
argue that both theory and experiment give evidence that the two twin
orientations are not necessarily equally abundant and that the ratio
of tunnelling currents in twinned and untwinned materials should be
related to the relative abundance of the two twin orientations.
\\{\tt preprint: cond-mat/9607125}
\end{abstract}


]

\section{Introduction}

The symmetry of the order parameter in the cuprates is the subject of
much current debate. While several experimental techniques such as
{\sc arpes}\cite{olson,shen,ding1,ding2} and {\sc
cits}\cite{edwards,edwards2} probe the magnitude of the
superconducting order parameter Josephson tunnelling is, perhaps, the
only observable phenomena which directly probes the phase of the
superconducting order parameter as well as, indirectly, its magnitude.

The focus of the debate has been upon whether the order parameter has
$s$-wave (even under a 90$^\circ$ rotation) or $d$-wave (odd under a
90$^\circ$ rotation) symmetry.  Since most experiments are performed
upon orthorhombic materials in which the $s$- and $d$-wave symmetries
belong to the same irreducible representation there is no clear
distinction between them and, rather than the symmetry of the order
parameter, the discussion should focus on whether there are nodes
present in the order parameter which cross the Fermi surface.

Although the $s$- and $d$-wave representations of the tetragonal
systems mix freely in orthorhombic systems we can still speak of the
``$s$'' and ``$d$'' parts of the order parameter if what we mean are,
respectively, the parts which are odd and even under a 90$^\circ$
rotation. It would, however, be highly unusual for one of these
components to be present in an orthorhombic superconductor without the
other being present as well. Further, if the odd, or $d$, part is
dominant then nodes which cross the Fermi surface will be present in
the order parameter (what we call a $(d+\epsilon s)$-wave order
parameter) while if the even, or $s$, part is dominant the nodes, if
they are present, will not cross the Fermi surface (what we call an
$(s+\epsilon d)$-wave order parameter). It is possible for an order
parameter that is even under a 90$^\circ$ rotation (ie, $s$-wave) to
have nodes but we consider this situation unlikely and we will not
examine this possibility here.

Several corner junction YBCO-Pb tunnelling
experiments\cite{mathai,brawner,wollman} are cited as strong evidence
for the order parameter in YBCO having $d$-wave type symmetry.  Other
tunnelling experiments which contain only YBCO in a bi- and
tri-crystal rings\cite{tsuei1,kirtley,tsuei2} also give strong
indications of order parameters with $d$-wave type symmetry.  Both of
these types of experiments have the plane of the junctions
perpendicular to the $a$-$b$ plane -- what we refer to as ``edge
junction'' experiments. It is the observation of $\pi$ phase shifts in
the corner junctions and half flux quanta in the ring experiments that
is the strong evidence for the order parameter in YBCO having $d$-wave
type symmetry.\cite{Hg} Note that for the observation of $\pi$ phase
shifts there is no difference between a $d$-wave order parameter and a
$(d+\epsilon s)$-wave order parameter.

The results of these edge junction experiments are independent of the
presence of twins in these materials.\cite{mathai-talk} This indicates
that the order parameter is phase-locked across the twin boundaries,
although calculations indicate\cite{andy,sigrist} that the magnitudes
(as well as the relative magnitudes and perhaps the relative phase as
well) of the $d$ and $s$ may change in the twin boundary. Although
this phase-locking is not unexpected it may have other consequences.

There is a second class of YBCO-Pb tunnelling experiments in which the
plane of the junction is parallel to the $a$-$b$ plane -- what we
refer to as ``$c$-axis junction'' experiments.\cite{sun,tanaka} The
presence of a current in these experiments is cited as evidence
against a $d$-wave order parameter in the
literature\cite{sigrist,elvis} as well as articles for more general
physics audiences.\cite{levi} In fact, due to the orthorhombic
symmetry of these materials the order parameter will have $(d+\epsilon
s)$-wave symmetry and a $c$-axis tunnelling current of reduced
magnitude is expected in untwinned
materials.\cite{odonovan3,odonovan10} The current is {\em not} just
due to the $\epsilon s$ part of the order parameter but is also caused
by the $d$ part of the order parameter due to the orthorhombicity of
the Fermi surface, a fact that appears not to be widely appreciated.

While there is quite a large variation in the maximum
resistance-tunnelling current product, $I_cR_N$, observed in these
$c$-axis tunnelling experiments, $I_cR_N$ in the untwinned samples is, on
average, about twice that in the twinned samples.\cite{dynes} The
argument that a finite $c$-axis YBCO-Pb tunnelling current is evidence
against a $d$-wave order parameter is based upon the intuitive
assumption that in the presence of twins any orthorhombicity in a
phase-locked order parameter will average out over the junction
area. In fact, theory clearly shows\cite{horovitz} that this would be
an unusual condition and that one twin should be significantly more
abundant that the other. An experiment which measures the relative
abundance of the two twins by comparing the intensities of the $(200)$
and $(020)$ diffraction peaks\cite{gaulin} indicates that one of the
two twins is, depending upon the specific sample, 2-3 times more
abundant that the other.\cite{gaulin}

If the ratio of the two possible twin orientations is $n$:$m$ and if
the current-resistance product, $I_cR_N(T=0)$, is $I_cR_N^{(\circ)}$ for an
untwinned junction then for the twinned sample the current
should be:
\begin{equation}
I_cR_N^{\rm\tiny (twin)}=\left|\frac{n-m}{n+m}\right|I_cR_N^{(\circ)}.
\end{equation}

This is consistent with the experimental observation of Sun {\it et
al.}\cite{dynes} in which the Josephson tunnelling current in
untwinned $c$-axis YBCO-Pb is about double that observed in twinned
junctions if the ratio of twins is approximately 3:1.

In Section II we present a {\sc bcs} formalism for a model system in
which a tetragonal CuO$_2$ plane layer is coupled to an orthorhombic
CuO chain layer and give the equation for the resistance-tunnelling
current product, $I_cR_N$. In section III we present the results of
some representative calculations as well as some experimental results
on the relative abundances of the two twin orientations. In section
IV we make some concluding statements.

\section{Formalism}

For a bilayer system (ie, $\alpha=1,2$) the coupled {\sc bcs}
equations are:\cite{odonovan10}
\begin{eqnarray}
\label{bcs.eq}
\mit\Delta_{{\bf k},1}&=& \frac{1}{\Omega}\sum_{{\bf q}}{\left(
V_{{\bf k},{\bf q},11}\chi_{{\bf q},1}
+V_{{\bf k},{\bf q},12}\chi_{{\bf q},2}
\right)} \nonumber \\
\mit\Delta_{{\bf k},2}&=& \frac{1}{\Omega}\sum_{{\bf q}}{\left(
V_{{\bf k},{\bf q},12}\chi_{{\bf q},1}
+ V_{{\bf k},{\bf q},22}\chi_{{\bf q},2}
\right)},
\end{eqnarray}
\noindent where we have taken $V_{{\bf k},{\bf q},12}=V_{{\bf k},{\bf
q},21}$ and:
\begin{eqnarray*}
\chi_{{\bf q},\alpha}
&\equiv& \left<a_{{\bf q}\uparrow,\alpha} a_{-{\bf q}\downarrow,\alpha}\right>\\
&=&\frac{\mit\Delta_{{\bf q},\alpha}}{2E_{{\bf q},\alpha}}
\tanh \left( \frac{E_{{\bf q},\alpha}}{2 k_{\rm B}T} \right),
\end{eqnarray*}
is the the pair susceptibility, with:
\begin{eqnarray*}
E_{{\bf k},\alpha}=\sqrt{
\varepsilon_{{\bf k},\alpha}^2+\mit\Delta_{{\bf k},\alpha}^2},
\end{eqnarray*}
where $\varepsilon_{{\bf k},\alpha}$ are the band energies
in the normal state. Note that we have taken $\left<a_{{\bf
q}\uparrow,1} a_{-{\bf q}\downarrow,2}\right>=0$, ie there is no
pairing between electrons in different planes although other workers\cite{kuboki} do not make this
assumption.

We note that this set of equations (\ref{bcs.eq}) is invariant under
substitution $\{\mit\Delta_{{\bf k},2},V_{{\bf k},{\bf
q},12}\}\rightarrow
\{-\mit\Delta_{{\bf k},2},-V_{{\bf k},{\bf q},12}\}$ which means that the
overall sign of $V_{{\bf k},{\bf q},12}$ only affects the relative
sign of the order parameters in the two layers and not their
magnitudes. This is interesting because it means that the effect on
$T_{\rm c}$ of having an interlayer interaction is independent of
whether this interaction is attractive or repulsive, although the
$c$-axis Josephson tunneling current still depends upon the relative
sign of the interlayer interaction.

The $c$-axis Josephson junction resistance-tunneling current
product, $I_cR_N(T)$, through a superconductor-insulator-superconductor
junction for incoherent $c$-axis tunneling is given by the
relation:\cite{ambegaokar}
\begin{eqnarray}
\label{joe.eq}
I_cR_N(T)=&&\frac{2\pi T}{N^{\rm (Pb)}(0) N^{\rm (YBCO)}(0) \pi^2}\nonumber \\
&& \times\sum_n { A^{\rm (Pb)}(\omega_n)A^{\rm (YBCO)}(\omega_n)},
\end{eqnarray}
where:
\begin{eqnarray}
\label{integrand.eq}
A^{(\cdot)}(\omega_n)\equiv\frac{1}{\Omega}\sum_{\bf k}
\frac{\mit\Delta^{(\cdot)}_{\bf k}}{(\varepsilon^{(\cdot)}_{\bf
k})^2+(\mit\Delta^{(\cdot)}_{\bf k})^2+(\omega_n )^2},
\end{eqnarray}
in which the superscript $(\cdot)$ indicates on which side of the junction the
dispersion and order parameter are on, the sum over $\omega_n\equiv\pi
T(2n-1)$ is for all Matsubara frequencies, $R$ is the resistance of
the junction and $N^{(\cdot)}(0)$ is the normal state electronic {\sc
dos} given by:
\begin{eqnarray}
        N^{(\cdot)}(\omega)&=&\frac{1}{\Omega}\sum_{\bf k}\delta
        (\varepsilon^{(\cdot)}_{\bf k}-\omega) \nonumber\\
        &=&\lim_{\mit\Gamma\to 0}\frac{1}{\pi\Omega}\sum_{\bf
        k}\frac{\mit\Gamma}{(\varepsilon^{(\cdot)}_{\bf
        k}-\omega)^2+\mit\Gamma^2}.
\label{dos.eq}
\end{eqnarray}

Since the {\sc dos} and $\mit\Delta^{(\rm Pb)}_{\bf k}$ for lead are
constant the sum in Eq.~\ref{integrand.eq} can be performed and
$A^{\rm (Pb)}(\omega_n)$ is given by:\cite{ambegaokar}
\begin{eqnarray*}
A^{\rm (Pb)}(\omega_n) =
\frac{\mit\Delta^{(\rm Pb)}}{\sqrt{(\mit\Delta^{(\rm Pb)})^2 + (\omega_n
)^2}}.
\end{eqnarray*}

If the tunneling were coherent the matrix element (which is
incorporated into $R$) would have a $({\bf k-k^\prime})$ dependence,
and the sums over ${\bf k}$-space wouldn't be separable.

\section{Results}

       We use the same band structure as in our previous work on the
penetration depth of a coupled chain-plane bilayer\cite{odonovan10} in
which we assume tetragonal symmetry for the electron dispersion in the
CuO$_2$ plane with first and second neighbour hopping, while the
chains are quasi-one dimensional with very different hopping
probabilities in the $x$- and $y$- directions (the chains are along
the $y$-direction).  For the pairing potential, $V_{{\bf k},{\bf
q},\alpha\beta}$, which appears in the coupled BCS equations
(\ref{bcs.eq}) and which determines the superconductivity, we use a
nearly antiferromagnetic Fermi liquid model with magnetic
susceptibility of the phenomenological form given by Millis, Monien
and Pines ({\sc mmp}):\cite{mmp}
\begin{eqnarray*}
V_{{\bf k},{\bf q},\alpha\beta}=g_{\alpha\beta}
\frac{-\chi_\circ}{1+\xi_\circ^2|{\bf k-q-Q}|^2},
\end{eqnarray*}
where $\xi_\circ$ is the magnetic coherence length and ${\bf Q}$ is
the commensurate vector $(\pi,\pi)$ in the 2-D Brillouin zone. Our
results are not specific to this interaction -- any interaction which
would result in a $d$-wave order parameter in the CuO$_2$ plane layer
will give a $(d+\epsilon s)$-wave order parameter when the CuO$_2$
planes are coupled to the CuO chain layers.

\begin{figure}
{\setlength\tabcolsep{0pt}
\begin{tabular}{c c c c c}
	\makebox[0.3\columnwidth][l]{\large (a)} & &
	\makebox[0.3\columnwidth][l]{\large (c)} & &
	\makebox[0.3\columnwidth][l]{\large (e)}\\
                \epsfxsize=0.3\columnwidth
                \epsfbox{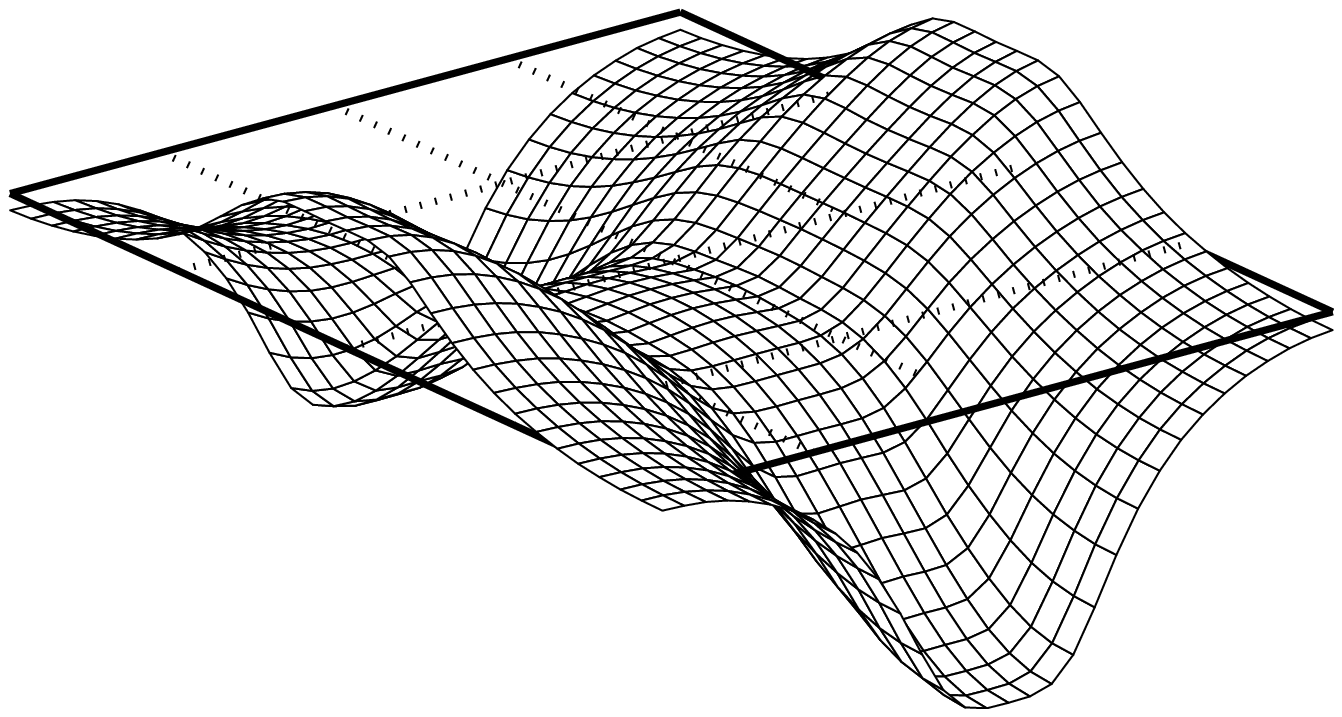} &
\parbox{10pt}{=\rule[-1.5cm]{0cm}{1.5cm}}&
                \epsfxsize=0.3\columnwidth
                \epsfbox{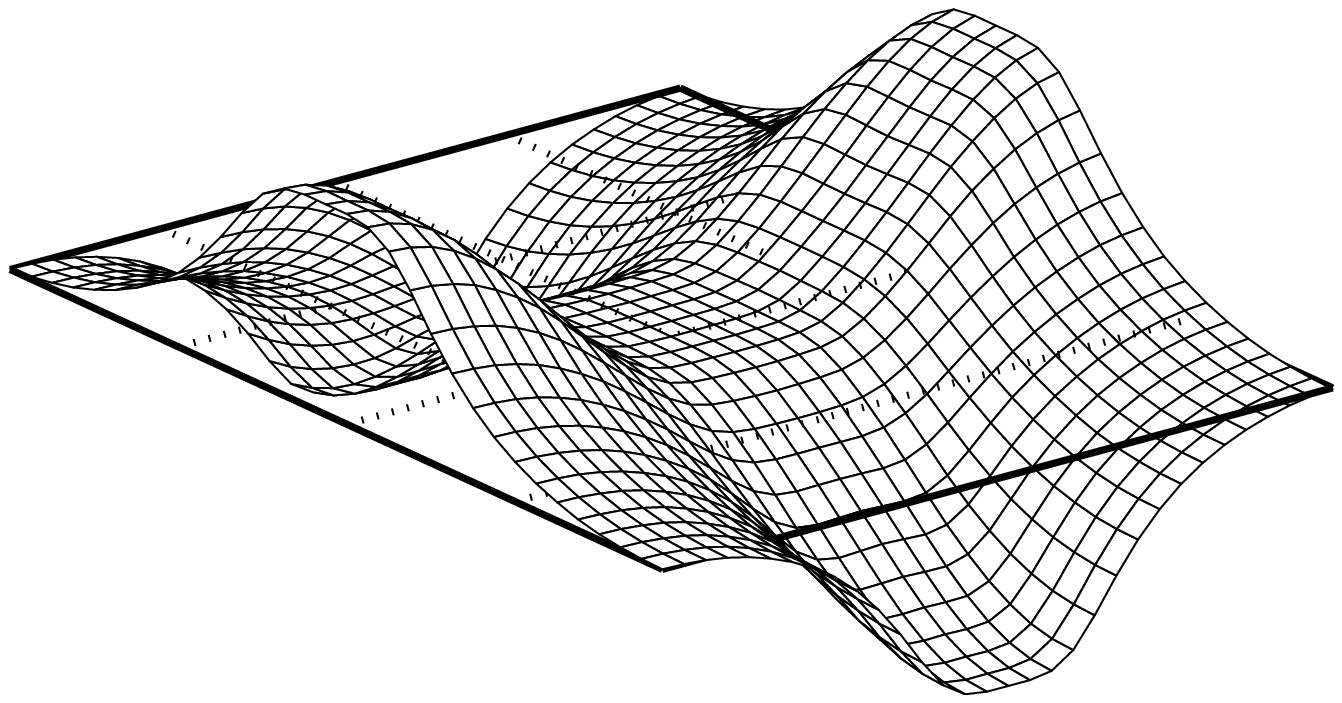} &
\parbox{10pt}{+\rule[-1.5cm]{0cm}{1.5cm}}&
                \epsfxsize=0.3\columnwidth
                \epsfbox{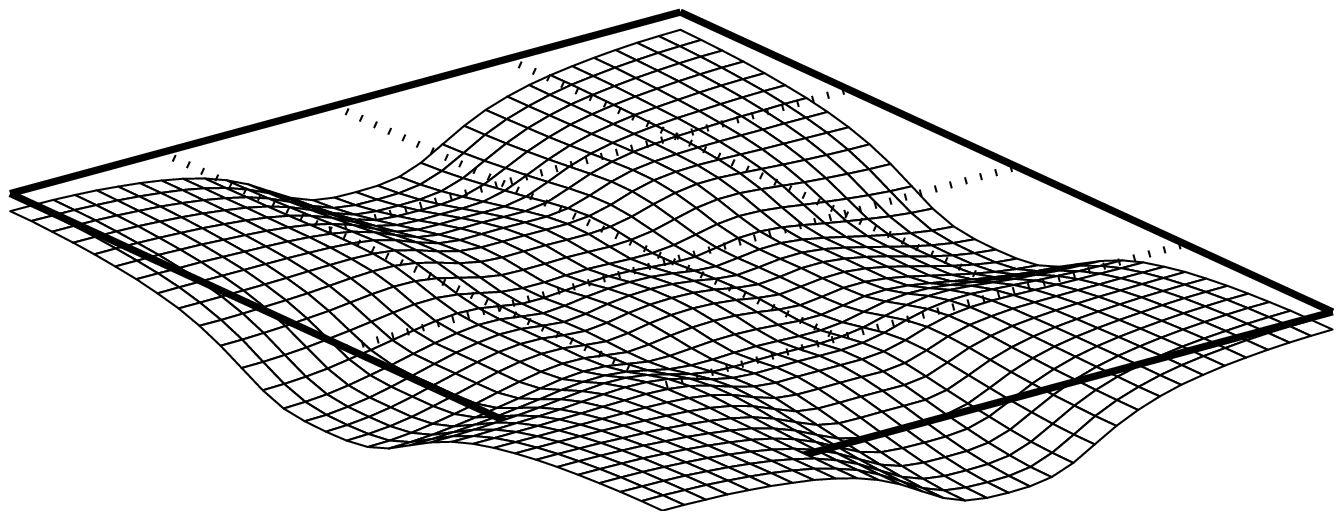} \\

	\makebox[0.3\columnwidth][l]{\large (b)} & &
	\makebox[0.3\columnwidth][l]{\large (d)} & &
	\makebox[0.3\columnwidth][l]{\large (f)} \\
                \epsfxsize=0.3\columnwidth
                \epsfbox{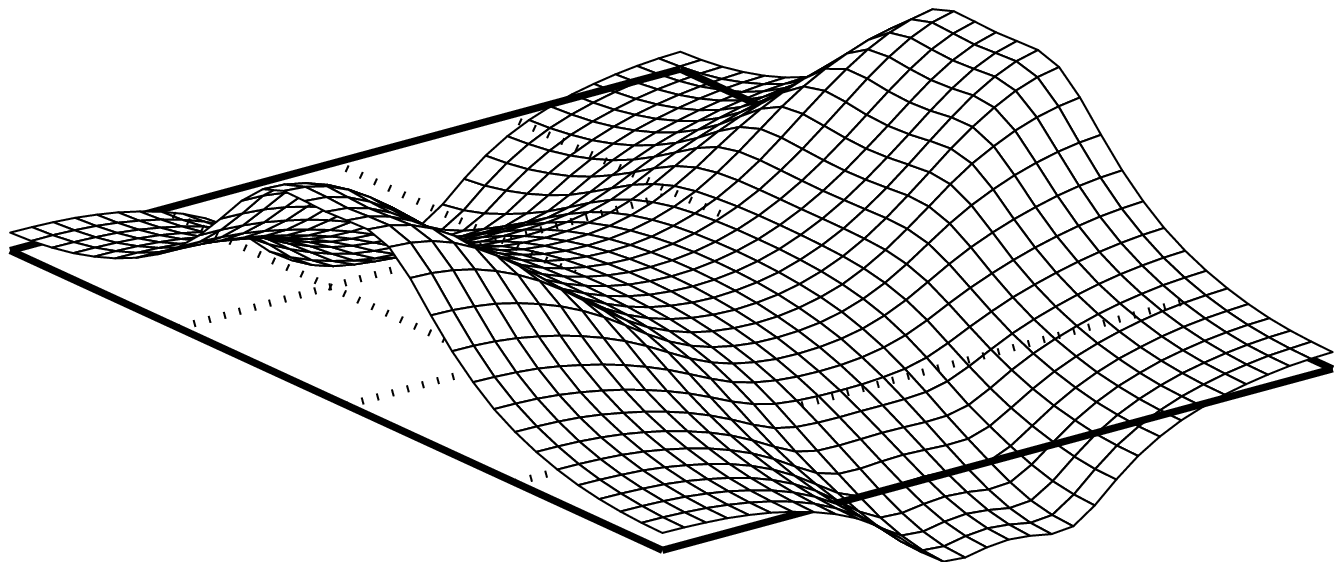} &
\parbox{10pt}{=\rule[-1.5cm]{0cm}{1.5cm}}&
                \epsfxsize=0.3\columnwidth
                \epsfbox{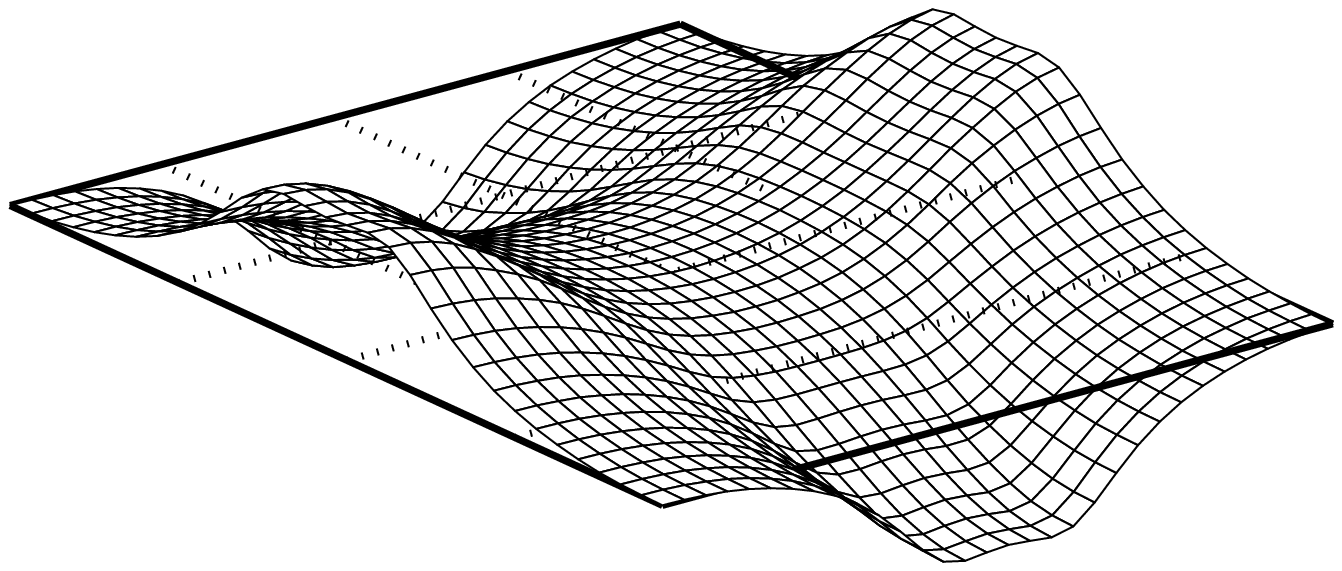} &
\parbox{10pt}{+\rule[-1.5cm]{0cm}{1.5cm}}&
                \epsfxsize=0.3\columnwidth
                \epsfbox{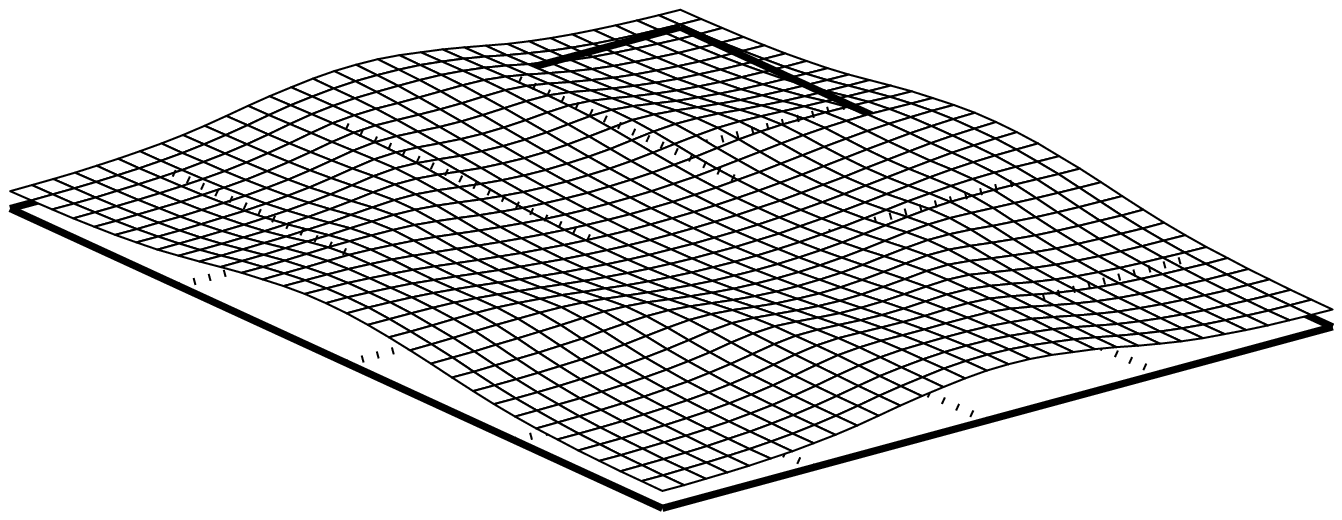}
\end{tabular}
}
\caption{
The order parameter in the first Brillouin zone for (a) the plane
layers and (b) the chain layers. Beside are the projections of the $d$
components, (c) and (d), and the $s$ components, (e) and (f). The
vertical scale in all frames is the same. Note that the relative phase
of the $d$-components in the two layers are the same while that of the
$s$-components are opposite; this is caused by the interlayer
interaction, $V_{{\bf k},{\bf q},12}$, being negative (ie,
repulsive).}
\label{gaps}
\end{figure}

No pairing interaction is assumed to act directly in the chain band,
ie $g_{22}=0$, so that the superconductivity in the chains is entirely
due to the $g_{12} = g_{21}\not =0$.  This parameter ($g_{12}$) is
fixed to get a critical temperature value of 100K for a chosen value
of the in plane pairing, $g_{11}$.  In Fig.~\ref{gaps}, we show
results for the gap value $\mit\Delta_{{\bf k},1}$ in the planes
(Fig.~\ref{gaps}a) and for $\mit\Delta_{{\bf k},2}$ in the chains
(Fig.~\ref{gaps}b) as a function of ${\bf k}$ in the first Brillouin
zone.  The order parameters result from a numerical solution of
equations (\ref{bcs.eq}).  For the runs shown in Fig.~\ref{gaps},
$\{g_{11},g_{12},g_{22}\}=\{26.2,10,0\}$ and $T_c=100$K.

       On the right hand side of Fig.\ref{gaps}, we have decomposed
the order parameters into $d$-wave (c) and (d) and $s$-wave (e) and
(f) components for the plane and chain, respectively.  While the main
component is certainly the $d$-wave part, the $s$-wave admixture is,
nevertheless, significant in magnitude in both the chain and plane
bands.

\begin{figure}[t]
	\begin{center}
	\begin{tabular}{c c c} 
			\makebox[0.3\columnwidth][l]{\large (a)} &
			\makebox[0.3\columnwidth][l]{\large (c)} &
			\makebox[0.3\columnwidth][l]{\large (e)} \\
			\epsfxsize=0.3\columnwidth
			\epsfbox{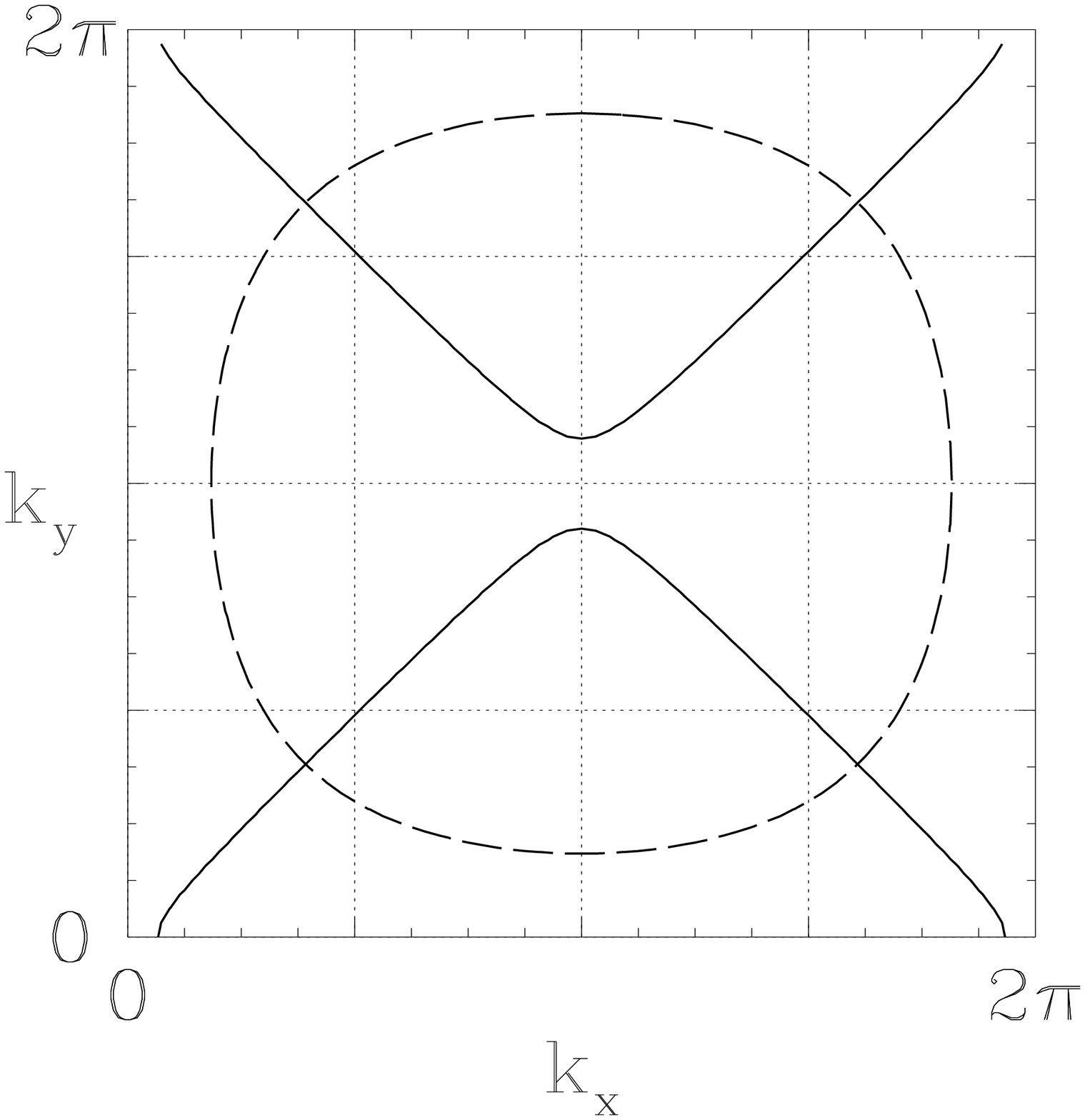} &
			\epsfxsize=0.3\columnwidth
			\epsfbox{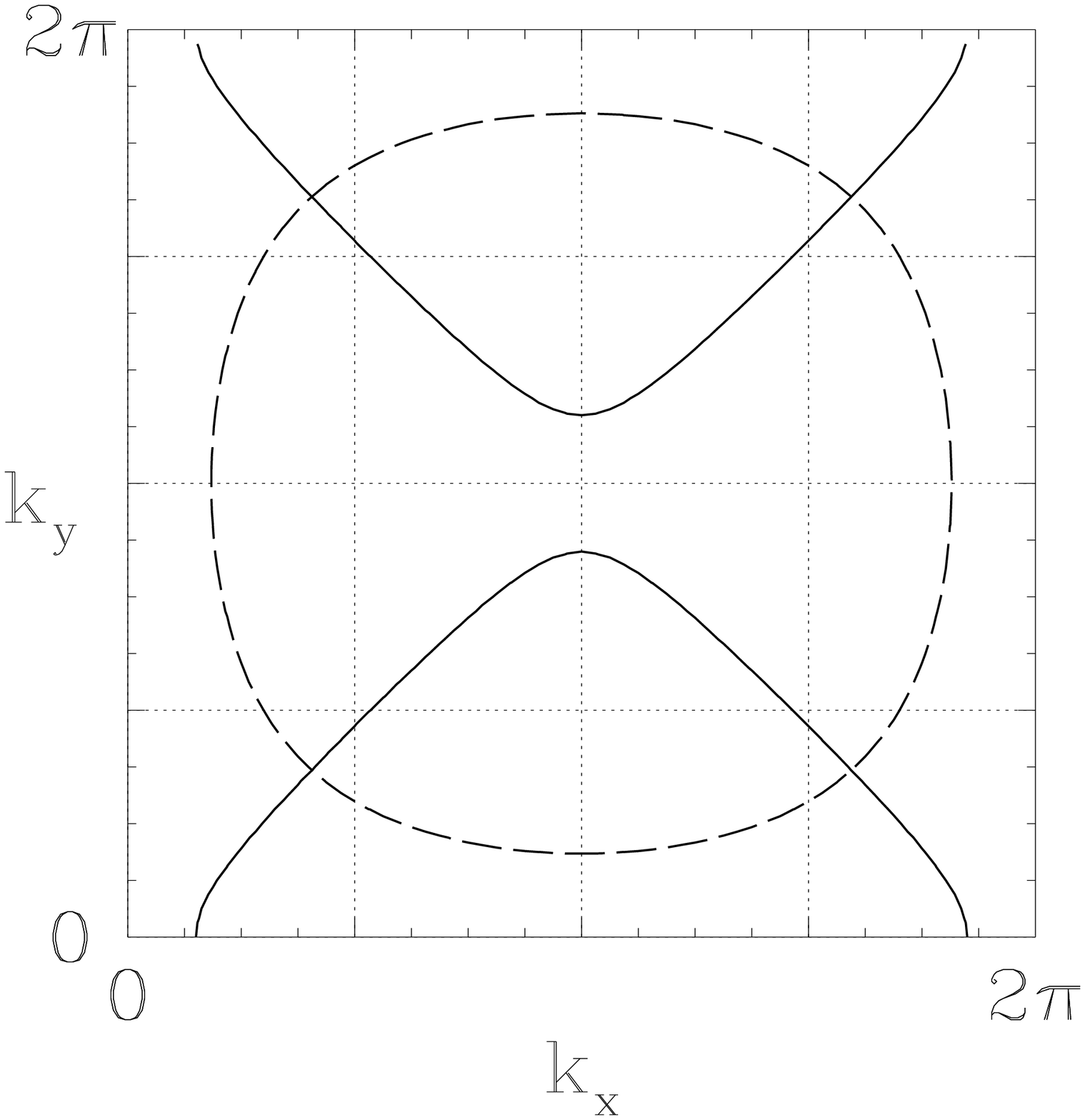} &
			\epsfxsize=0.3\columnwidth
			\epsfbox{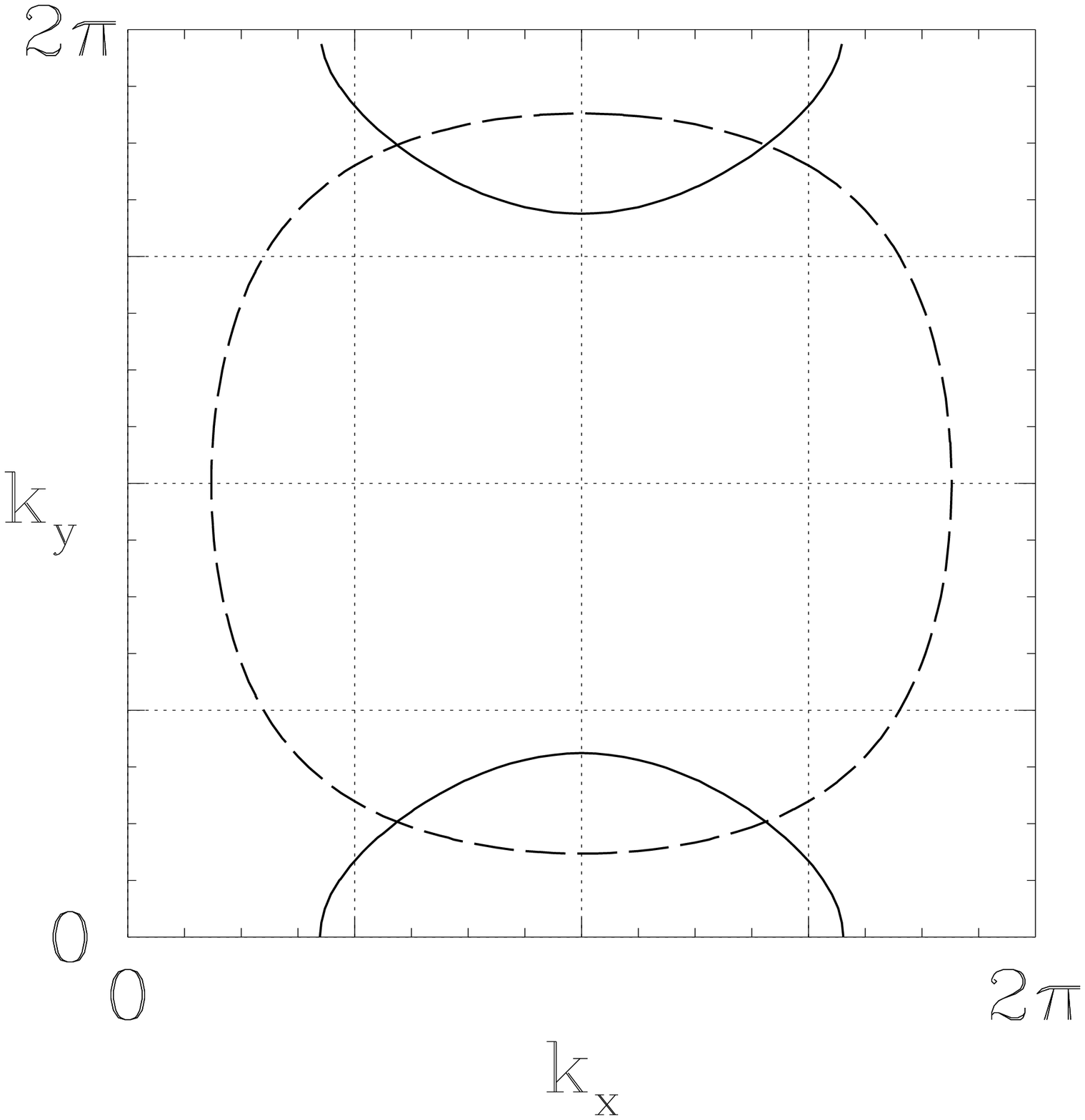} \\ 
			\makebox[0.3\columnwidth][l]{\large (b)} &
			\makebox[0.3\columnwidth][l]{\large (d)} &
			\makebox[0.3\columnwidth][l]{\large (f)} \\
			\epsfxsize=0.3\columnwidth
			\epsfbox{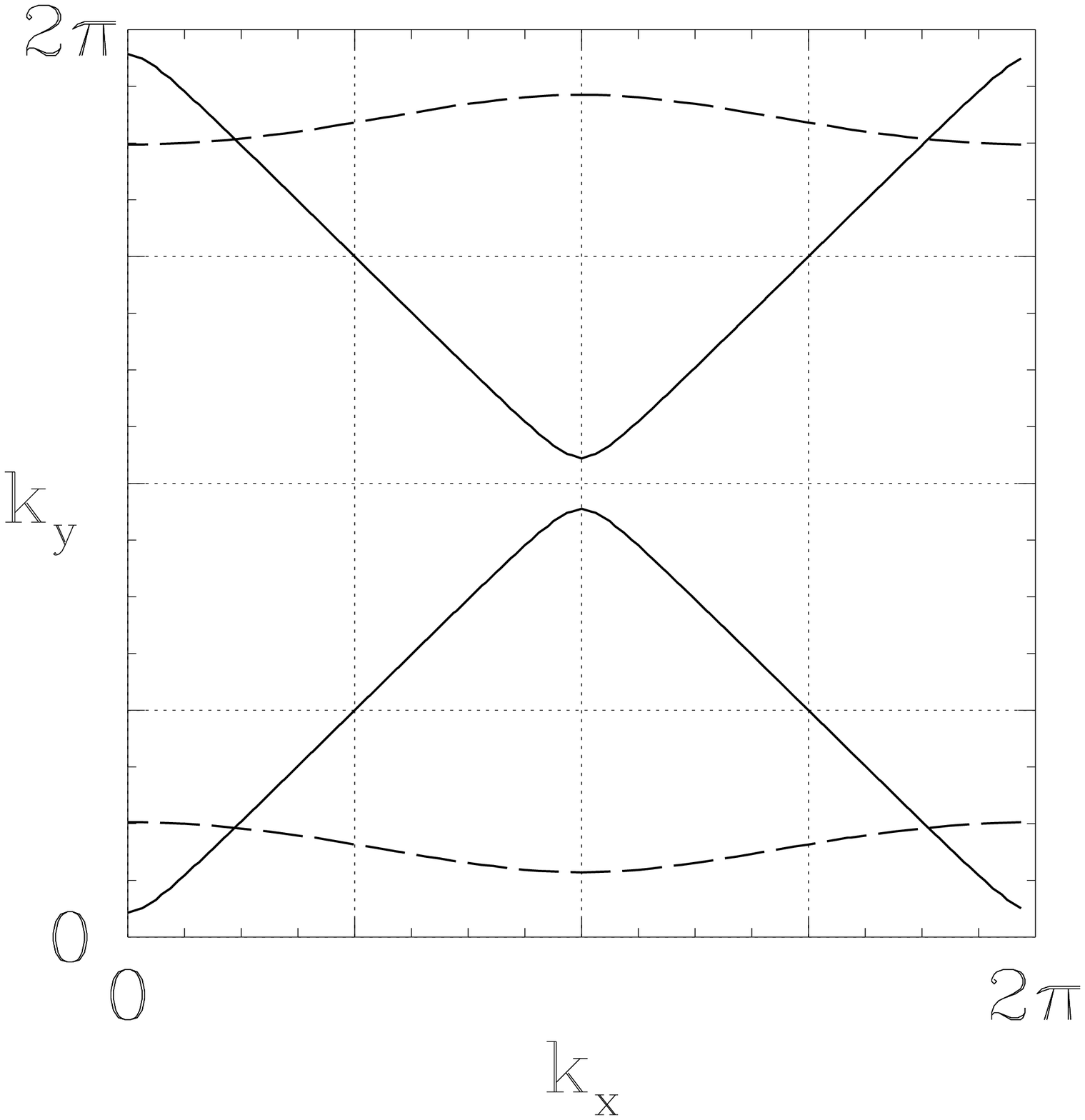} &
			\epsfxsize=0.3\columnwidth
			\epsfbox{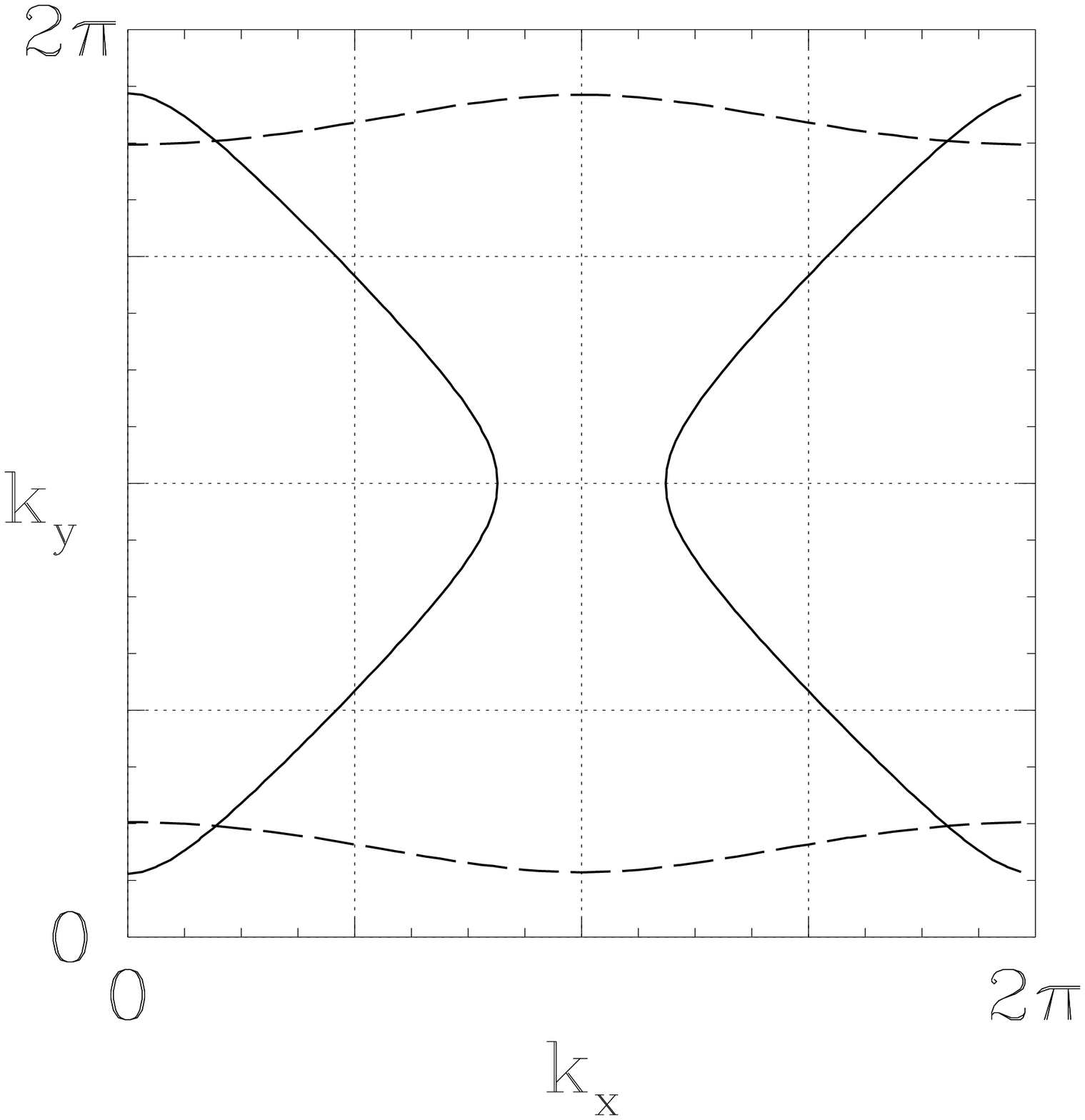} &
			\epsfxsize=0.3\columnwidth
			\epsfbox{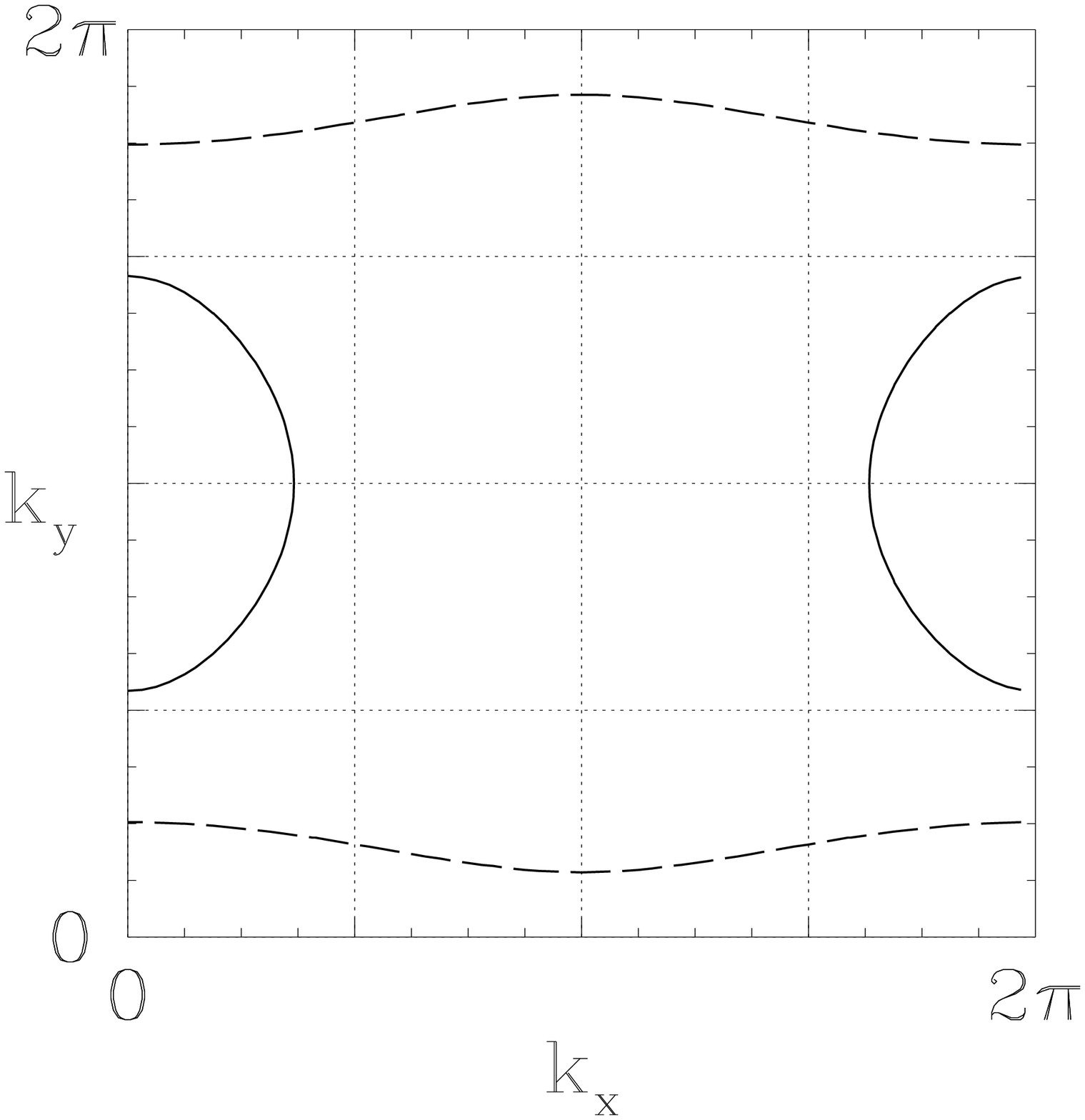}
		\end{tabular}
	\end{center}
\caption{
The Fermi surface (dashed curves) and gap nodes (solid curves) for a
CuO$_2$ plane layer (top frames) and a CuO chain layer (bottom frames)
for three different interlayer interaction strengths (left, middle and
right frames). As the strength of the interlayer interaction,
$g_{12}$, is increased (left to right frames) the proportion of the
$s$-component in both layers increases. If the interlayer interaction
were further increased the gap nodes would leave the Brillouin zone
altogether and the order parameter would become $s$-like. Note that
the Fermi surface in the CuO$_2$ plane layer is tetragonal but that
the gap node is orthorhombic.  }
\label{nodes}
\end{figure}

       A useful representation of these gap results is to show the
contours of gap zeros on the same plot as the Fermi surface.  This is
presented in the series of frames shown in Fig.~\ref{nodes}.  The top
frames apply to the plane while the bottom frames apply to the chains.
In all cases, (a), (c), (e) for the planes and (b), (d), (f) for the
chains, the same Fermi surface (dashed curves) was used.  By choice,
the Fermi contour have tetragonal symmetry in the top figure while the
chain Fermi surface is a quasi straight line along $k_x$ as is
expected for chains along the $y$-direction in configuration space.
The pictures are for three different values of pairing potential.  The
first set of two frames (a) and (b) are for
$\{g_{11},g_{12},g_{22}\}=\{29.9, 5, 0\}$, i.e. very little coupling
between chains and planes (off diagonal $g_{12}$ small).  In this
case, the gap in the plane is nearly pure $d$-wave as is also the
induced gap in the chains. The zero gap contours are given by the
solid line and would be the main diagonals in a pure $d$-wave
system. As the coupling $g_{12}$ is increased
$\{g_{11},g_{12},g_{22}\}=\{26.2, 10, 0\}$, a significant $s$-wave
component gets mixed into both solutions and the gap nodes move off
the main diagonals of the Brillouin zone (This is the solution that is
plotted in Fig.~\ref{gaps}).  The gap nodes still cross the Fermi
surfaces in both chains and planes.  As the coupling is further
increased to $\{g_{11},g_{12},g_{22}\}=\{9.18, 20, 0\}$,
Fig.~\ref{nodes} (e) and (d), the gap nodes move far off the diagonal
and for the chains they no longer cross the Fermi surface so that
there is a finite minimum value of the gap on this sheet of the Fermi
surface.

       Calculation of the product of Josephson current times
resistivity given in equation (\ref{joe.eq}) for a YBCO-I-Pb tunnel
junction gives for the intermediate case
$\{g_{11},g_{12},g_{22}\}=\{26.2, 10, 0\}$ the tunnelling current,
$I_cR_N(T=0)$, is 0.2-0.3 meV for the plane layers and $\pm$(2-3) meV
for the chain layers.  The relative sign depends upon whether the
interlayer interaction, $g_{12}$, is attractive or repulsive.  These
values for chain and plane tunnelling current, $I_cR_N$, are upper and
lower bounds and note that the expected value should be somewhere in
between since the $I_cR_N(T=0)$ values are dependent upon the amount
of tunnel junction area which is covering the uppermost chain or plane
layer.

\begin{figure}[t]
	\begin{center}
	\begin{tabular}{c c}
			\makebox[0.5\columnwidth][l]{\large (a)} &
			\makebox[0.5\columnwidth][l]{\large (b)} \\
			\epsfxsize=0.5\columnwidth
			\epsfbox{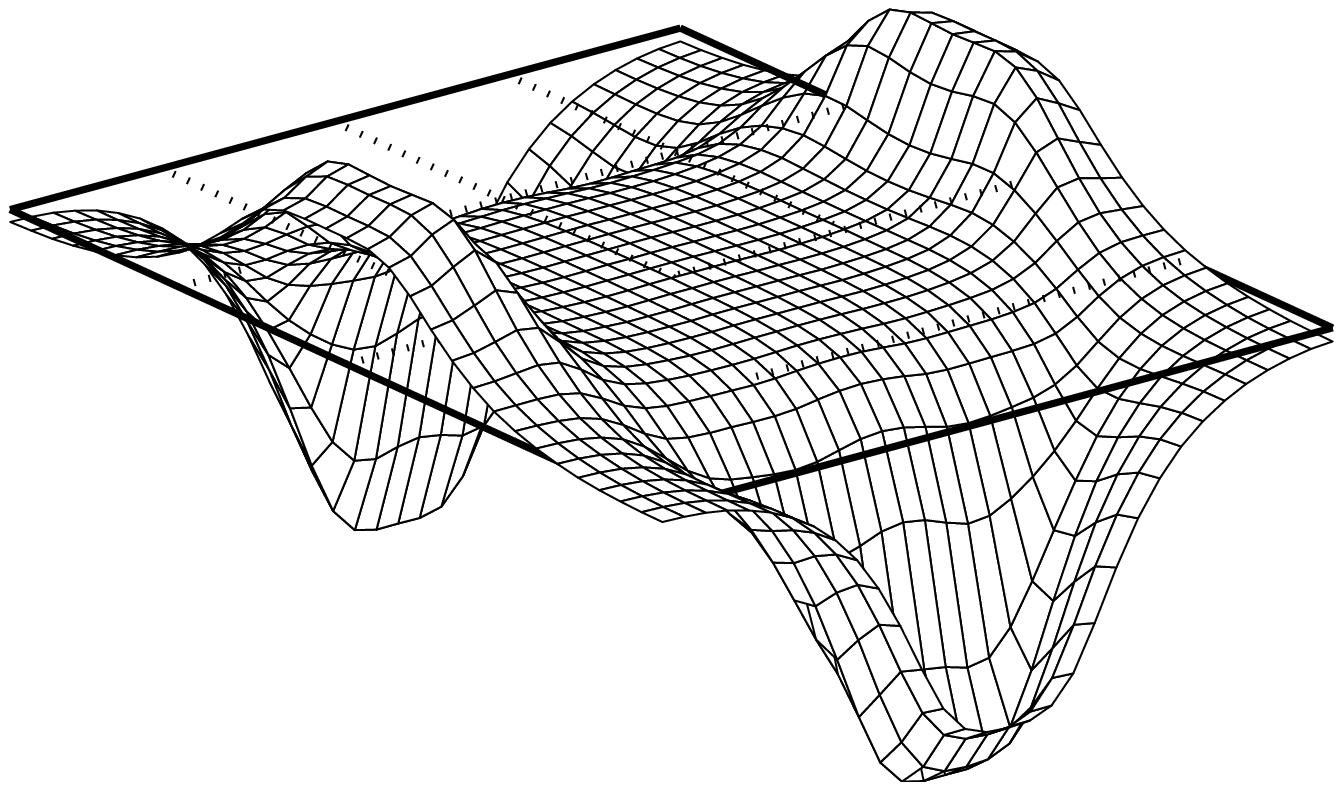} &
			\epsfxsize=0.5\columnwidth
			\epsfbox{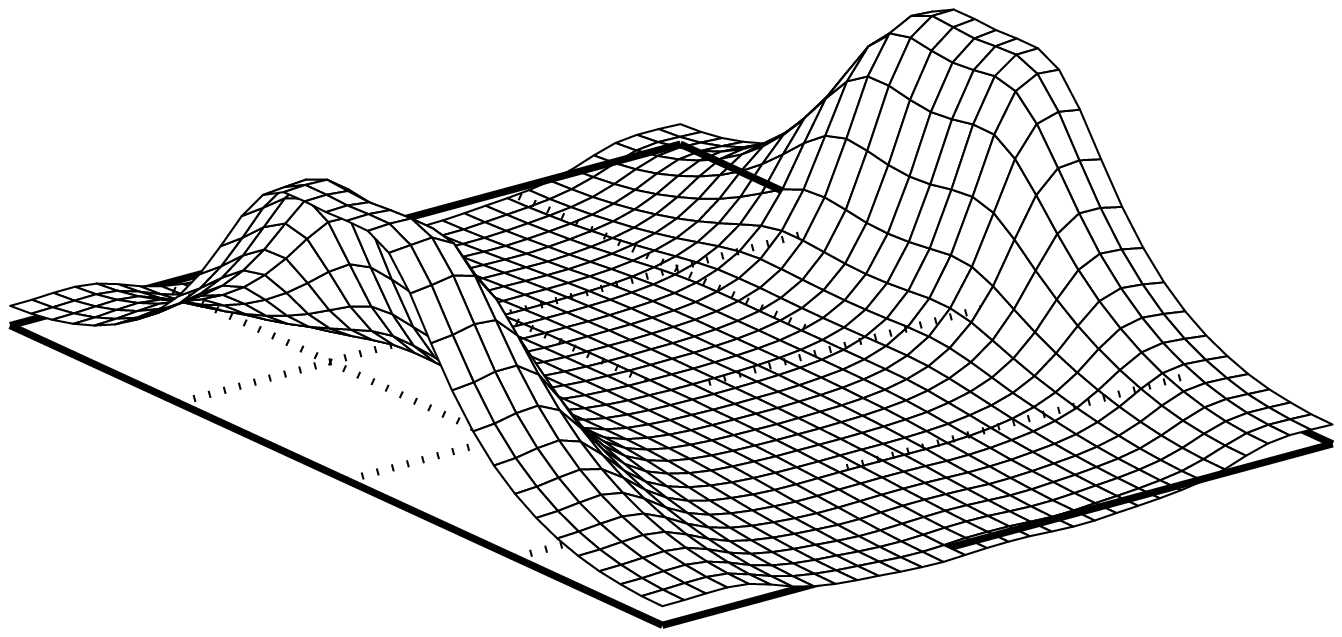}
		\end{tabular}
	\end{center}
\caption{
The ${\bf k}$-space integrand of $A^{\rm (YBCO)}(\omega_n)$ (see
Eq.~\protect\ref{joe.eq}) for (a) the CuO$_2$ plane layer and (b) the
CuO chain layer. In the plane layer the positive and negative parts
mostly cancel and the resulting current is small while in the chain
layer only a very small amount of the integrand, $A^{\rm
(YBCO)}(\omega_n)$, is negative and the resulting current is
large. This effect is due to the different Fermi surfaces in the two
layers.  }
\label{integrand.fig}
\end{figure}

       The order of magnitude difference obtained between plane and
chain layer for the $I_cR_N(T=0)$ product is understood as follows.  For a
given Matsubara frequency, $\omega_n$ in equation
(\ref{integrand.eq}), the contribution to the sum over ${\bf k}$,
which ranges over the entire Brillouin zone, is strongly peaked about
the Fermi surface because the denominator in (\ref{integrand.eq})
becomes smallest in this case.  This is seen clearly in
Fig.~\ref{integrand.fig} where we show the integrand of equation
(\ref{integrand.eq}), again for the intermedicate case, as a function
of ${\bf k}$ in the first Brillouin zone for planes (a) and chains
(b), respectively for a particular Matsubara frequency,
$\omega_n=$50meV.  It is clear that in the plane layer the positive
and negative parts largely cancel each other.  They would, in fact,
give exactly zero if the gap had pure $d$-wave symmetry.  The
situation for the chains is completely different because the Fermi
surface now does not have tetragonal symmetry and even if the gap was
pure $d$-wave there would not be a large degree of cancellation
between the positive and negative regions.  {\em Thus, we note the
important result that the main part of the Josephson current coming
from the chains is due to the d-wave part of the gap function and
would still be large if we did not account for the $s$-wave admixture.
Thus, in an experiment on an untwinned single crystal of YBCO the
Josephson current coming from the chain part of the Fermi surface is
sampling mainly the $d$-wave part of the gap and therefore, such
experiments do not reflect directly the $s$-wave admixture.}

       It has been argued that for twinned samples, the $c$-axis
Josephson current will cancel because of the cancellation between $a$-
and $b$-twins.\cite{dynes} This argument would apply equally well to
our work, since we should then average over pictures as shown in
Fig.\ref{integrand.fig}(b) with opposite phases ($a$- and $b$-twins).
However, we do not expect twins to be present in equal numbers, and
the expected cancellation will not occur.\cite{horovitz}

High resolution x-ray scattering measurements have been carried out on
a small single crystal of YBa$_{2}$Cu$_{3}$O$_{6.93}$ for the purpose
of investigating its twin structure.  Measurements were made with an
18kW rotating anode x-ray source and a double-axis diffractometer
using a perfect Ge(111) monochromator with sufficient resolution to
cleanly separate Cu K$_{\alpha1}$ from Cu K$_{\alpha2}$ radiation.
Full details will be given in a separate publication.\cite{gaulin}

\begin{figure}[t]
	\begin{center}
	\begin{tabular}{c c c c}
			\epsfxsize=\columnwidth
			\epsfbox{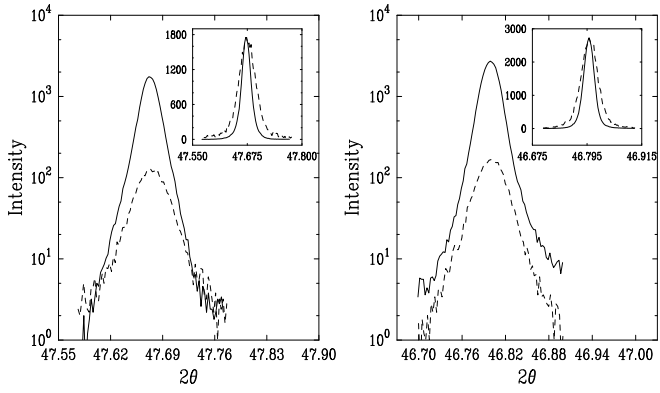} \\
		\end{tabular}
	\end{center}
\caption{
Representative longitudinal x-ray scattering scans through the (2,0,0)
and (0,2,0) Bragg peaks of the YBa$_{2}$Cu$_{3}$O$_{6.93}$ single
crystal are shown on a logarithmic scale.  It is clear that the peak
intensity of the majority phase twin domains are at least an order of
magnitude stronger than that of the corresponding minority phase
domain.  The inset shows the same data, but now on a linear intensity
axis, and scaled so that the peak intensities for the two domains
coincide.  The Bragg features of the minority twin domains are clearly
much broader than those of the majority phase domains.  }
\label{fig4}
\end{figure}

The crystal under study was grown by the UBC group and had approximate
dimensions of 1$\times$1$\times$0.03 mm$^3$, with the c-axis oriented
along the thin dimension.  The crystal was sufficiently thin that
scattering within the orthorhombic basal plane (such as (2,0,0) and
(0,2,0)) could be performed in transmission geometry and such
measurements thus probe the bulk of the crystal.  Representative
longitudinal scans of the (2,0,0) and (0,2,0) Bragg peaks are shown
for both twin domains in Figure 4.  As is clear from this figure, one
twin domain has a peak intensity which is more than an order of
magnitude stronger than the other.  This leads naturally to an
interpretation of the twin structure in the crystal in terms of a
majority and a minority domain.  Also, as would be expected in this
scenario, the minority phase lineshape is very noticeably broader than
that of the majority phase, whose lineshape appears to be
approximately resolution limited.  This is shown in the inset to
Figure~\ref{fig4}, where the same data is plotted on a linear
intensity axis, and has been scaled so that the peak intensity of the
two domains coincide.

The majority and minority phase domain distribution in this crystal
was found to be inhomogeneous.  X-ray scattering measurements were
performed with a very narrow ($\sim$0.05 mm) incident beam to allow
measurements which probe different regions of the crystal.  The twin
structure was investigated as the sample was translated through the
narrow beam along the majority phase a-axis.  This investigation
yielded results which ranged from completely untwinned on one extreme
edge of the crystal to an approximate 1 to 1 ratio of the volume
fraction of majority to minority domains on the other edge.  Interior
regions of the crystal yielded some intermediate value of this ratio.
We estimate the average ratio of the volume fraction occupied by the
majority and minority twin domains for the entire sample to be about
2-3 to 1.  While these measurements reveal a complex inhomogeneous
morphology to the twin structure, it is certainly clear that
macroscopic regions of the crystal exist in which one domain
predominates over the other.

\section{Conclusions}

       For $c$-axis incoherent YBCO-I-Pb Josephson tunnelling
junctions, the $d$-wave component of the gap parameter can contribute
very significantly because of the orthorhombic nature of the chain
Fermi surface which emphasizes  the contributions to the Josephson
current from those parts of the Brillouin zone along the Fermi
surface.  The $s$-component of the gap in the chains and planes will
also contribute but this may be less important so that such
experiments on untwinned samples do not probe directly the $s$-wave
admixture of a predominantly $d$-wave gap function.

       We argue that the observation of a $c$-axis tunnelling current
in the experiments of Sun {\it et al.}\cite{sun} and Tanaka {\it et
al.}\cite{tanaka} on twinned samples can also be understood within our
theory because we do not expect equal numbers of $a$- and $b$-twins in
their experiments.

\section*{Acknowledgements}

Research supported in part by the Natural Sciences and Engineering
Research Council of Canada ({\sc nserc}), the Ontario Centre for
Materials Research ({\sc ocmr}) and by the Canadian Institute for
Advanced Research ({\sc ciar}). We would like to thank Bob Goodings
for discussions and insights and D. Bonn for providing the crystal.

\end{document}